\newcommand{\mb}[1]{\ifmmode#1\else\mbox{$#1$}\fi}

\newcommand\al{\mb{\alpha}}

\newcommand\be{\mb{\beta}}
\newcommand\ga{\mb{\gamma}}
\newcommand\de{\mb{\delta}}

\newcommand\la{\mb{\lambda}}
\newcommand\si{\mb{\sigma}}

\newcommand\Ga{\mb{\Gamma}}

\newcommand\Om{\mb{\Omega}}

\newcommand{\beq}{\begin{equation}}
\newcommand{\eeq}{\end{equation}}

\newcommand{\bea}{\begin{eqnarray}}
\newcommand{\eea}{\end{eqnarray}}
\newcommand{\mod}[1]{\mid \! {#1} \! \mid}

\newcommand{\deriv}[2]{\frac{d {#1}}{d {#2}}}

\newcommand{\x}{\mb{\times}}

\newcommand{\corr}[1]{\langle {#1} \rangle}
\newcommand{\gsim}
{\raise.3ex\hbox{$\;>$\kern-.75em\lower1ex\hbox{$\sim$}$\:$}}
\newcommand{\lsim}
{\raise.3ex\hbox{$\;<$\kern-.75em\lower1ex\hbox{$\sim$}$\:$}}

\documentstyle[prl,aps,twocolumn,floats]{revtex}

\begin{document}

%%%%%%%%%%%%       abstract and title page    %%%%%%%%%%%%%%%%%%%%%%

%\preprint{}
\twocolumn[\hsize\textwidth\columnwidth\hsize\csname @twocolumnfalse\endcsname
\title{Cosmological Birefringence and the Microwave Background}
\author{Nathan\  F.\  Lepora}
\address{King's College, Cambridge University, Cambridge, CB2 1ST,
England} 
\date{\today}
\maketitle

\begin{abstract}
We show that significant anisotropy in electromagnetic propagation
generates a distinctive signature in the microwave background. 
The anisotropy may be determined by looking at the cross correlator
of the $E$-mode and $B$-mode polarisation spectrum.\\
\end{abstract}
]

%%%%%%%%%%%%%         the text   %%%%%%%%%%%%%%%%%%%%%%%%%%%%%%%%%%%

Linearly polarised light travels through a birefringent Universe the
same as it would through a birefringent crystal; the angle of
polarisation rotating dependant upon its direction of
propagation. As such it represents a deviation from the Cosmological
principle --- where the large scale Universe is postulated to be
homogenous and equivalent in all directions --- with the Universe
possessing a fundamental spatial anisotropy attributed to the
properties of electromagnetic propagation.   

Theoretically, probably the best motivation for such an anisotropy
would be a small homogenous background torsion component to
the background geometry of the Universe, as pointed out by Dobado and
Maroto~\cite{doba}. Torsion couples into the Dirac equation through
minimal coupling in the covariant 
derivative. Regularisation of the full electromagnetic-fermion
Lagrangian then produces an effective birefringence for
propagating electromagnetic radiation~\cite{doba}. This birefringence
takes the form of a dipole for the rotation of polarisation
\beq
\label{dipole}
\be(\mbox{{\bf r}}) =  \frac{d}{2\Ga_s} \hat{\mbox{\bf r}} \cdot
\hat{\mbox{\bf s}},  
\eeq                               
from polarised light emitted at light distance $d$, direction
$\hat{\mbox{\bf r}}$. The axis $\hat{\mbox{\bf s}}$ and length scale
$\Ga_s$ are related to the spatial component of the con-torsion vector
$S_\al = \epsilon_{\al \be \ga \de} T^{\be \ga \de}$ by  
\beq 
\Ga_s^{-1} \hat{\mbox{\bf s}} = - \frac{\al Q^2}{6\pi} {\mbox{\bf S}},
\eeq
with $\al$ the fine structure constant, $\hat{\mbox{\bf s}}$ a unit
vector and $Q^2$ the total squared fermionic charge. 
For the Standard Model $Q^2=8$; other models lead to different values,
for examples the supersymmetric standard model has $Q^2=12$. On a
heuristic level it is not surprising that torsion gives 
this effect because of the association with rotation, first described
by Cartan in 1922~\cite{cart}. 

Interest in birefringence has been revived recently because
of Nodland and Ralston's claim to have measured a Hubble distance scale
dipolar birefringence 
in the synchrotron radiation of radio galaxies~\cite{ral1}. They claim
to have obtained, to about $0.1\%$ significance, $\Ga_s \sim 0.1
cH_0^{-1}$ and $\hat{\mbox{\bf s}}=(0^\circ 
\pm 20^\circ {\rm decl}, 21 \pm 2 hrs {\rm R.A.})$. Curiously, within
the limits of experimental error $\hat{\mbox{\bf s}}$ coincides with
both the dipole axis 
of the microwave background~\cite{Brace} and the rotation axes of
galaxies in the Perseus-Pisces super-cluster\cite{kuhn}. We should
point out that historically this claim predates Dobado and Morato's
paper, which provided a theoretical explanation with a con-torsion of
magnitude about $10^{-30}$eV.

Their analysis has received some criticism~\cite{crit1,crit2}, which
they in turn refute~\cite{refu}. The major criticism, made by several
independent groups, is that their choice of statistics was
intrinsically biased towards producing birefringence~\cite{crit2}. 
Different statistics indicate significant birefringence is not
supported for length scales $\Ga_s \lsim 0.4 cH_0^{-1}$.

The purpose of this letter is to point out that if large scale
birefringence exists then the effects of it should be present in
the radiation of 
the microwave background. In the near future it should be possible
look for this effect, which should present a very distinctive
signature onto the microwave background polarisation spectrum.

Essentially the point is that the temperature variations of the
microwave background originate in the metric perturbations within the
surface of last scattering. These metric perturbations also cause
potential flows of matter, which gives rise to local partial
polarisations of the microwave background by Thompson
scattering. These polarisations split into contributions respectively
uncorrelated and correlated with the temperature fluctuations, with
the correlated component forming distinctive patterns around the
temperature fluctuations~\cite{neil}. However, if there is appreciable
anisotropy in the polarisation propagation then
the orientation of our measured polarisations will be skewed relative
to the temperature fluctuation map. That is the basis of this letter.

Specifically, we shall discuss the effects of birefringence on {\em
scalar} perturbations, before briefly mentioning the effect on vector
and tensor modes at the end of this letter. Before discussing the
effects it is necessary to discuss the polarisation spectra in more
depth. For more details we refer the reader to the excellent review
article by Hu and White~\cite{hu97}.

Microwave background polarisation is induced by the last
Thompson scattering of a decoupling photon. The cross section 
depends upon the incident and scattered polarisations 
$\hat{\mbox{\bf e}}$ and $\hat{\mbox{\bf e}}'$ as
\beq
\deriv{\si_T}{\Omega} \propto \mod{\hat{\mbox{\bf e}} \cdot
\hat{\mbox{\bf e}}'}^2,
\eeq
peaking for perpendicular scattering with parallel incident and
scattered polarisations. The photon field is tightly bound to the
baryons, forming a net photon-baryon fluid, with the incident photon
field depending upon the potential flow within the last scattering
surface of last scattering. Symmetry means that local isotropic or
dipolar flows produce no net effect, since scattered polarisations
cancel. Thus the lowest order non-trivial local moment of the flow is
quadrupolar, producing linear polarisation directed along the
compressional axis of the quadrupole.

Consider a rotationally symmetric perturbation within the last
scattering surface. If the perturbation represents a potential well,
corresponding to a temperature cold spot, the photon-baryon fluid flows 
radially inwards giving azimuthally orientated local quadrupole
moments; this produces an azimuthal polarisation
pattern. Alternatively, for potential hill perturbations,
corresponding to hot spots, the quadrupoles are rotated though ninety
degrees producing a radial polarisation pattern.
Considering the magnitude of polarisation, single valuedness forces
the polarisation to vanish in the centre of the perturbation, with the
magnitude increasing radially outwards then falling off as the
gradient flow decreases. One should note that the two polarisations
are orientated respectively parallel and normal to the direction of
maximal polarisation gradient. This is a generic feature of scalar
potentials. 

The above argument is, of course, an idealisation and generally one
expects the measured polarisations be dominated by random
fluctuations. These fluctuations obscure the temperature correlated
component. Statistically, however, one still expects some correlation
with the temperature anisotropies at a level of about
$15\%$\cite{neil}. 

Now consider a temperature-polarisation map of the
microwave background obtained directly at the surface of last
scattering for those photon that will propagate to
us. Temperature represents a scalar function  
distributed across the surface of last scattering $T(\theta, \phi)$,
whilst polarisation is a vector function ${\mbox{\bf P}}(\theta,
\phi)$ such that ${\mbox{\bf P}} \cdot \hat{\mbox{\bf r}}=0$. Because
polarisation is 
${\mbox{\bf C}}P^1$ valued $\pm {\mbox{\bf P}}$ are identified. As
mentioned above only about $15\%$ of this polarisation represents a
component correlated with the temperature.

The photon field then propagates to us, red shifting as the Universe
expands. If contaminatory effects, such as obscuration by
our galaxy, effects of intervening galaxies and reionisation are
successfully subtracted, then there are left two modifying features of
the homogenous large scale geometry :\\
(i) Curvature of the intervening Universe will effect the angular
scale of perturbations in $T$ and ${\mbox{\bf P}}$. However, this
curvature will be determined by the effects on the acoustic peak
spectrum, and may be consistently taken into account.\\
(ii) Birefringence, if present, will alter the polarisation
spectrum. Generally, the magnitude of the polarisation field
$\mod{\mbox{\bf P}}$ will be unaffected, whilst the direction will
be rotated by an angle $\be(\theta, \phi)$; the measured polarisation
will then take the form
\beq
\label{pol-bir}
{\mbox{\bf P}}_{\rm bir}(\theta, \phi) =
\left( \begin{array}{cc} 
\cos\be & -\sin\be \\ 
\sin\be & \cos\be \end{array} \right){\mbox{\bf P}}.
\eeq
If birefringence originates via a small homogeneous torsion,
and we choose spherical polar coordinates such that the axis
$\theta=0$ coincides with the con-torsion, then the birefringence
measure takes the form $\be(\hat{\mbox{\bf r}})= (d_{\rm
rec}/2\Ga_s)\cos \theta$. Here $d_{\rm rec}$ is the light distance to
the 
surface of last scattering taken to good approximation to be the
current horizon distance
\beq
d_{\rm rec}(\Om_0) = \left\{
\begin{array}{cc}
\frac{1}{\sqrt{\Om_0-1}}\cos^{-1}\left(\frac{2}{\Om_0}-1\right)
cH_0^{-1} & \Om_0>1\\
2cH_0^{-1} & \Om_0=1\\
\frac{1}{\sqrt{1-\Om_0}}\cosh^{-1}\left(\frac{2}{\Om_0}-1\right)
cH_0^{-1} & \Om_0<1
\end{array} \right.
\eeq

When measuring the polarisation of the microwave background it is
convenient to express it as components of the local parity
eigenstates. These correspond to the $E$-mode, the $E>0$ ($E<0$)
component perpendicular (parallel) to the maximal gradient of
$\mod{\mbox{\bf P}}$; and the $B$-mode, at $45^\circ$ to the $E$-mode
component. These are parity eigenstates 
because if one imagines a homogenous distribution of $E$ or $B$-mode
polarisation on a sphere then the configurations are respectively
parity even and parity odd. A basis of $E$-mode and $B$-mode
polarisation vectors is particularly useful because it neatly
separates out perturbations with particular parity signatures: for 
scalar perturbations parity forces the polarisations {\em to be purely
$E$-mode}. This observation is the basis of the following
discussion.

The point is that birefringence converts $E$-mode polarisation into
$B$-mode polarisation. $E$-mode polarisations within the last
scattering surface, $P^E_{\rm l.s.}$, from different regions on the
sky will experience conversion into both $B$-mode and $E$-mode
polarisations such that 
\beq
\label{polrot}
\left( \begin{array}{c} P^E \\ P^B \end{array} \right) =
\left( \begin{array}{cc} 
\cos2\be  & -\sin2\be \\ \sin2\be & \cos2\be \end{array} \right)
\left( \begin{array}{c} P^E_{\rm l.s.} \\ 0 \end{array} \right) ,
\eeq
with $\be(\theta, \phi)$ representing the amount of birefringent
rotation experienced by a photon emitted from $(\theta, \phi)$ on the
sky. The angle $2\be$ originates from the non-orthogonality of the
polarisation basis.

Assuming $\be(\theta, \phi)$ does not vary substantially on a small
scale, say $\la^\circ$, then we may bin the sky intoareas of
about $\la^\circ \x \la^\circ$ with approximately constant $\be$
within each. Denoting the $n$th such region by $(\theta_n, \phi_n)$
and the correlator over that region by $\corr{\cdot \cdot}_n$,
Eq.~(\ref{polrot}) gives the $EB$ cross correlator on the region $n$
to be 
\beq
\corr{P^E P^B}_n = -\frac{1}{2}\sin4\be(\theta_n, \phi_n)
\corr{P^E_{\rm l.s.}P^E_{\rm l.s.}}_n.
\eeq
Assuming all polarisation is from scalar modes,
$\corr{P^E_{\rm l.s.}P^E_{\rm l.s.}}_n = \corr{P^2}_n$; then the
birefringence on a scale $\la^\circ$ is
\beq
\be(\theta_n, \phi_n)=-\frac{1}{4} \sin^{-1}\left( 2\frac{\corr{P^E
P^B}_n}{\corr{P^2}}
\right). 
\eeq
Both correlators $\corr{P^EP^B}_n$ and $\corr{P^2}_n$ are determinable
from a polarisation map of sufficient accuracy and resolution, with
\beq
\corr{P^2}_n = \corr{P^EP^E}_n + \corr{P^BP^B}_n +
\sqrt{2}\corr{P^EP^B}_n.
\eeq 
From an experimental point of view it is probably best to 
pick $\la$ to be about the largest scale at which $\be$ is
approximately constant to obtain the best statistics. We should also
mention that the above relation does not require an all sky coverage
and good determinations could be made from a local region of the sky.

Consider again birefringence originating from a small homogenous
con-torsion field. For sufficient birefringence one would see
alternating parallel stripes of 
$E$-mode and $B$-mode polarisations forming concentric circles on the
sky, with the centres coincident in the direction of con-torsion. On
stripes of maximal $E$-mode the $B$-mode is negligible and {\em
vice-versa}. Between stripes the polarisation continuously  
interpolates between one mode and the other. The number of stripes,
depends upon the magnitude of the birefringence, and a simple
calculation gives $n$ stripes for $\Ga_s \approx 2d_{\rm rec}/n$. For
the birefringence obtained by Nodland and Ralston this translates to
$20$ stripes in a critical density Universe. For the lower bounds on
birefringence obtained by other authors one obtains $n \lsim 5$. 

For a sufficiently small con-torsion such that $\Ga_s \gsim 2d_{\rm
rec}$, a pattern of stripes is not obtained. Instead perpendicular to
the con-torsion the polarisation is purely $E$-mode,
interpolating to a mixture of $E$-mode and $B$-mode polarisation at
the poles.  

In addition to producing significant $E$-$B$ cross correlation,
birefringence should also affect the $E$-mode polarisation multipole
spectrum. The polarisation anisotropy spectrum is described by the
multipole expansion of the two point correlation function
\bea
C^E(\vartheta)
&=& \corr{P^E(\hat{\mbox{\bf r}}) P^E(\hat{\mbox{\bf
r}}')}_{\hat{\mbox{\bf r}} \cdot \hat{\mbox{\bf r}}' = \cos
\vartheta}\\ 
&=& \frac{1}{4\pi} \sum_l (2l+1) C_l^E P_l(\cos \vartheta), 
\eea
with the coefficients $C_l^E$ extractable from observations. An
analogous expression represents $C^B(\vartheta)$ in terms of its
coefficients $C_l^B$. Representing the initial coefficients at the last
scattering surface by $(C^E_l)_{\rm l.s.}$,
Eq.~(\ref{polrot}) implies that the effect of birefringence on the
multipole expansion is to perform a linear transformation upon the
coefficients. After birefringence present day coefficients are
linearly related to the coefficients at last scattering by 
\bea
\label{ce}
C^E_l &=& A_{lm} (C^E_l)_{\rm l.s.},\\
\label{cb}
C^B_l &=& B_{lm} (C^E_l)_{\rm l.s.},
\eea
with
\bea
\label{m}
A_{lm} &=& -\frac{1}{2\pi N_{lm}} \int_{\varphi=0}^{2\pi}
\int_{\vartheta=0}^{\pi} P_l(\cos \vartheta) P_m(\cos \vartheta) 
\cos 2\be(\vartheta,\varphi) \sin \vartheta 
{\rm d}\vartheta {\rm d}\varphi,\\ 
\label{n}
B_{lm} &=& -\frac{1}{2\pi N_{lm}} \int_{\varphi=0}^{2\pi}
\int_{\vartheta=0}^{\pi} P_l(\cos \vartheta) P_m(\cos \vartheta) 
\sin 2\be(\vartheta,\varphi) \sin \vartheta 
{\rm d}\vartheta {\rm d}\varphi,
\eea
and normalisation $N_{lm}=\sqrt{(2l+1)(2m+1)}$.
Relations~(\ref{ce}, \ref{cb}) and (\ref{m}, \ref{n}) extract the last
scattering polarisation from the measured polarisation.  Were
birefringence to be detected, comparison to the cosmological model
prediction for $(C^E_l)_{\rm l.s.}$ would offer a useful consistency 
check on the results.  

Again we illustrate the above with a small homogenous con-torsion. In
this case the linear transformations $A_{lm}$ and $B_{lm}$ take the
form 
\bea
A_{lm} &=& -\frac{1}{N_{lm}}
\int_{x=-1}^{1} P_l(x) P_m(x) \cos nx {\rm d}x,\\
B_{lm} &=& -\frac{1}{N_{lm}}
\int_{x=-1}^{1} P_l(x) P_m(x) \sin nx {\rm d}x,
\eea
with $n \approx 2d_{\rm rec}/\Ga_s$ assumed to of the order one or
greater. These lead to modification of the polarisation spectrum
after birefringence, which conveniently splits into two
contributions. Firstly, correlation of the stripes produces a new
peak in the correlation function at scales $l \sim n$. Secondly,
rotation of $E$-mode into $B$-mode should approximately half the total
power in the $E$-mode polarisation spectrum. However, one should note
that a new low $l$ peak would be obscured by the reionisation peak,
produced by reionisation of the Universe at red shifts $z \sim 5-20$.

Summing up, significant birefringence of a magnitude below the current
experimental bounds would lead to a distinctive modification of the
microwave background polarisation spectrum. Its signature would
be conversion of $E$-mode polarisation into $B$-mode polarisation,
with a pattern relating to the form of the birefringence. Some
modification of the $E$-mode power spectrum would also occur, 
offering a useful consistency check on such an effect. If birefringence
is significant in the post photon-baryon coupled Universe then its
effect will be seen in the next generation of microwave background
experiments. 

We finish on a few points that warrant further note:\\
(i) Although we have concentrated specifically on illustrating
birefringence with a model of homogenous torsion, it is likely that
other reasonable models should lead to similar consequences. 
Particularly compelling is a model with an axionic condensate whose
density varies linearly across the horizon~\cite{pri-neil}, as
predicted in some texture models. Such a model would give rise to a
dipolar birefringence. Additionally the matter gradient would orientate
the birefringence axis with the dipole axis of the microwave
background.\\  
\pagebreak 
(ii) Birefringence would also effect any vector or tensor
perturbations. Vector perturbations produce mainly $B$-mode
polarisations, so birefringence would convert this into $E$-mode
polarisation. Tensor perturbations, such as those from gravitational
waves, give a mixture of $E$ and $B$-mode polarisation; birefringence
should mix these contributions. However, it is generally expected
that scalar perturbations dominate the density spectrum.\\ 
(iii) We have assumed that for scalar potentials the initial
polarisations are purely $E$-mode. Torsion would alter
the parity properties of the potential, producing
some initial $B$-mode polarisation. However, this contribution should
be small since the comparitive length scales between birefringence and
the perturbations differ by several orders of magnitude.  \\ 
(iv) Although we have presented torsion as a reasonable theoretical
motivation for birefringence, the absence of detectable birefringence
in the microwave background would constitute an upper bound upon
torsion for cosmologically relevant distance scales. The expected
sensitivity should be around $10^{-32}$eV; several orders of magnitude
better than local torsion estimates, whose best current upper
bound limits the local torsion to less than
$10^{-18}$eV~\cite{lamm}.\\ 
(v) It should be mentioned that it would be difficult to reconcile a
significant torsion with a conventional spin density source. Assuming
that particles have spin approximately $\hbar$, a torsion around
$10^{-31}$eV would require a number density about $10^{40}$cm$^{-3}$,
far in 
excess of any conventional particle densities. Observation of a
relevant birefringence would necessitate some rethinking of how
torsion is sourced. 

%%%%%%%%%%%%%%%%%%%%%%%%%%%%% acknowledgements %%%%%%%%%%%%%%%%%%%

\begin{acknowledgements}
I acknowledge King's College, Cambridge, for a junior research
fellowship and thank P. Saffin, N. Turok and J. Weller for their
valuable advice relating to this work.  
\end{acknowledgements}

%%%%%%%%%%%%%%%%%%%%%%%%%%%  bibliography   %%%%%%%%%%%%%%%%%%%%%%%


\begin{thebibliography}{99}

\bibitem{doba}
A. Dobado and A. Maroto,
{\em Mod. Phys. Lett. A} {\bf 12}, 3003 (1997).

\bibitem{cart}
E. Cartan,
{\em Compt. Rend. Acad. Sci.} {\bf 174}, 593 (1922).

\bibitem{ral1}
B. Nodland and J. P. Ralston, 
{\em Phys. Rev. Lett.} {\bf 78}, 3043 (1997).

\bibitem{Brace}
R. Bracewell and V. Eshleman, preprint aps1997jun13-006 (1997).

\bibitem{kuhn}
R. Kuhne,
{\em Mod. Phys. Lett. A} {\bf 12}, 2473 (1997).

\bibitem{crit1}
J. P. Leahy, astro-ph/9704285;
J. F. C. Wardle, R. A. Perley and M. H. Cohen,
{\em Phys. Rev. Lett.} {\bf 79}, 1801 (1997).

\bibitem{crit2}
D. J. Eisenstein and E. F. Bunn,
{\em Phys. Rev. Lett.} {\bf 79}, 1957 (1997);
S. M. Carroll and G. B. Field,
{\em Phys. Rev. Lett.} {\bf 79}, 2394 (1997);
T. J. Loredo, E. E. Flanagan and I. M. Wasserman,
{\em Phys. Rev. D} {\bf 56}, 7507 (1997).

\bibitem{refu}
B. Nodland and J. P. Ralston, astro-ph/9706126;
B. Nodland and J. P. Ralston, 
{\em Phys. Rev. Lett.} {\bf 79}, 1958 (1997).

\bibitem{neil}
D. Coulson, R. G. Crittenden, and N. G. Turok,
{\em Phys. Rev. Lett.} {\bf 73}, 2390 (1994).

\bibitem{hu97}
W. Hu and M. White,
{\em New Astron.} {\bf 2}, 323 (1997).

\bibitem{pri-neil}
N. Turok, private communication.

\bibitem{lamm}
C. L\"ammerzahl, gr-qc/9704047.

\end{thebibliography}
\end{document}